\documentclass[nologo,11pt,a4paper,hidelinks]{ETHpaper}
\usepackage{graphicx, epsfig, amsmath, amssymb,color,wasysym}
\usepackage[numbers]{natbib}
\usepackage{subcaption}
\usepackage{booktabs}
\usepackage{mathtools}
\usepackage{cancel}
\usepackage[table]{xcolor}
\usepackage{colortbl}
\usepackage{algorithm2e}
\usepackage{dcolumn} 
\definecolor{lgray}{RGB}{210,210,210}
\newcolumntype{d}[1]{D{.}{.}{#1}}

\begin{document}

\newcommand{\mean}[1]{\left\langle #1 \right\rangle} 
\newcommand{\abs}[1]{\left| #1 \right|}

\title{Control contribution identifies \\ top driver nodes in complex networks}   

\titlealternative{Control contribution identifies top driver nodes in complex networks}   

\author{Yan Zhang, Antonios Garas, Frank Schweitzer}

\authoralternative{Yan Zhang, Antonios Garas, Frank Schweitzer}

\address{Chair of Systems Design, ETH Zurich, Weinbergstrasse 58, 8092, Zurich, Switzerland}

\reference{ACS - Advances in Complex Systems (submitted)}
\www{\url{http://www.sg.ethz.ch}}

\makeframing
\maketitle

\begin{abstract}

  We propose a new measure to quantify the impact of a node $i$ in controlling a directed network.
  This measure, called ``control contribution'' $\mathcal{C}_{i}$, combines the probability for node $i$ to appear in a set of driver nodes and the probability for other nodes to be controlled by $i$.
  To calculate $\mathcal{C}_{i}$, we propose an optimization method based on random samples of minimum sets of drivers.
  Using real-world and synthetic networks, we find very broad distributions of $C_{i}$.
    Ranking nodes according to their $C_{i}$ values allows us to identify the top driver nodes that control most of the network.
  We show that this ranking is superior to rankings based on control capacity or control range.
  We find that control contribution indeed contains new information that cannot be traced back to degree, control capacity or control range of a node.
  \end{abstract}

\section{Introduction}

System design implies the ability to steer the dynamics of a system such that a desired system state is reached. 
One way of achieving this goal is rooted in control theory, which utilizes components of the system as drivers for the dynamics \cite{Luenberger1979}.
This relates to the fundamental question if and how a system can be controlled \citep{Liu2016}.
It requires to have a proper system representation.
Here we use a complex network approach, in which system elements are represented by nodes and their interactions by links.
\citet{Liu2011} introduced an analytical framework called \emph{``network controllability''} that combines network theory and classical control theory. 
This framework has important applications for the study of empirical  systems \cite{Nacher2013c,Yuan2014,Srihari2013,Liu2014b,LX2019,Wuchty2014a,Vinayagam2016}, if interactions between system elements are captured by means of a large-scale complex network topology. 

Network controllability assumes that there is a dynamics \emph{on the network}, specifically that the state of nodes can be described by a linear dynamics.
Then, the minimum set of driver nodes required to control the whole system can be efficiently identified using this framework.
Given that not every node necessarily is a driver node, it is important to quantify to what extent a node contributes to controlling the whole network.
To solve this problem is the aim of our paper.
We want to rank driver nodes according to their contribution to controlling the network, in particular we want to identify the \emph{top drivers} with the largest contribution.

In the literature, various measures have been introduced to quantify the \emph{topological importance} of nodes.
Examples include  degree centrality,  PageRank,  coreness \cite{Kitsak2010,Garas2012b} and betweenness centrality \cite{Freeman1977}.
These measures, however, do not help us in solving our problem because they do not consider a dynamics on the network.
There are, indeed, other measures that take this dynamics into account, for instance \emph{control range}, $\mathcal{R}$, \cite{Wang2012e} and \emph{control capacity}, $\mathcal{K}$ \cite{Jia2013f}.
$\mathcal{R}$ quantifies the size of the subnetwork, i.e. the number of nodes, controlled by one driver node, and $\mathcal{K}$  quantifies the likelihood that a node is a driver node.
In Sect. \ref{sec:ident-sets-driv}, we will give examples for that. 
These measures, however, separate individual aspects of drivers, therefore it is not clear which measure should be effectively used to identify top driver nodes.   

In this work, we propose a new measure to identify top drivers, called \emph{``control contribution''}, $\mathcal{C}$, which is a \emph{node property}. 
Intuitively, the control contribution $\mathcal{C}_{i}$ of node $i$ captures the probability for any node in a network to be controlled by node $i$ joint with the probability that $i$ becomes a driver. 
To calculate $\mathcal{C}_{i}$, we propose an optimization method based on random samples of minimum sets of drivers that will be explained in Sect. \ref{sec:control-contribution}.
Calculating the distribution $P(\mathcal{C})$ for different real-world and synthetic networks, we find that the distribution is always very broad and does not follow a specific pattern.
Looking into relations to \emph{topological} quantities such as degree distribution,
we find that the degree distribution does not determine the distribution of control contributions. 
Further, no uniform pattern in the correlation exists between control capacity $\mathcal{K}$ and control range $\mathcal{R}$ that determine control contribution. 
Therefore, we argue that control contribution $\mathcal{C}$ indeed contains new information. 

We demonstrate that driver nodes chosen according to their $\mathcal{C}_{i}$ values lead to a larger part of a network that can be controlled. 
In this respect, nodes with a higher $\mathcal{C}_{i}$ outperform nodes chosen according to their $\mathcal{K}_{i}$ or $\mathcal{R}_{i}$ values. 
Therefore, our new quantity \emph{control contribution}, $\mathcal{C}_{i}$, efficiently identifies the top driver nodes in a network.

\section{Control contribution}
\label{sec:control-contribution}

\subsection{Identifying sets of driver nodes}
\label{sec:ident-sets-driv}

To introduce the concept of control contribution, we consider a directed network of $N$ nodes.
In this network, each node $i$ is captured by a state variable $x_{i}(t)$, and the states of all nodes can be described by the state vector $\mathbf{X}(t)=\{x_{1}(t),x_{2}(t),...,x_{N}(t)\}$ with $\mathbf{X} \in \mathbb{R}^N$.
Some of these nodes are controlled directly by the control signals $u_{k}(t)$.
We call these nodes \emph{drivers} and $N_{c}$ the number of driver nodes.
We denote $\mathbf{U}(t)  \in \mathbb{R}^{N_{c}}$ as the vector of control signals.
The matrix $\mathbf{B} \in \mathbb{R}^{N \times N_c}$ maps these signals to driver nodes, $b_{ij} \neq 0$ when control signal $j$ is attached to node $i$.

We further assume that $\mathbf{X}(t)$ follows the linear dynamics:  
\begin{equation} \label{eq1}
\dot{\mathbf{X}}(t)=\mathbf{AX}(t)+\mathbf{BU}(t), 
\end{equation}
with time-invariant matrices $\mathbf{A}$. $\mathbf{B}$ and $\mathbf{U}$.
$\mathbf{A} \in \mathbb{R}^{N \times N}$ is the interaction matrix with elements $a_{ij}$ $(i,j=1,...,N)$ describing the strength in which node $j$ can influence node $i$.
According to the the Kalman rank condition~\cite{kalman1963}, the linear system defined by Eq. \eqref{eq1} is controllable, if and only if the controllability matrix $\mathbf{C}=[\mathbf{B,AB,A^2B,...,A^{N-1}B}] \in \mathbb{R}^{N \times (N\cdot N_c)}$ has full rank, i.e., rank$(\mathbf{C})$=$N$. 

In many cases, we do not have the precise value of the non-zero elements in $\mathbf{A}$ and $\mathbf{B}$, therefore, it is not feasible to calculate rank($\mathbf{C}$) to check for controllability.
For these cases, we have the weaker requirement of structural controllability~\cite{Lin1974}. 
The key idea is to treat both the adjacency matrix $\mathbf{A}$ and the mapping matrix $\mathbf{B}$ as \emph{structural matrices} whose non-zero elements are free parameters.
Then the system is controllable iff the free parameters can be chosen such that rank$(\mathbf{C})$=$N$.

In general, rank$(\mathbf{C})$ denotes the controllable subsystem size, $N_{b}$, i.e. the number of nodes in the network that can be controlled.
$N_{b}$ not always equals $N$, but is smaller, which means the system is partially controllable.
We then define the fraction $n_{b}=N_{b}/N$ as the relative size of the controllable system.

\begin{figure}[t]
	\centering
  \includegraphics[width=0.95\columnwidth]{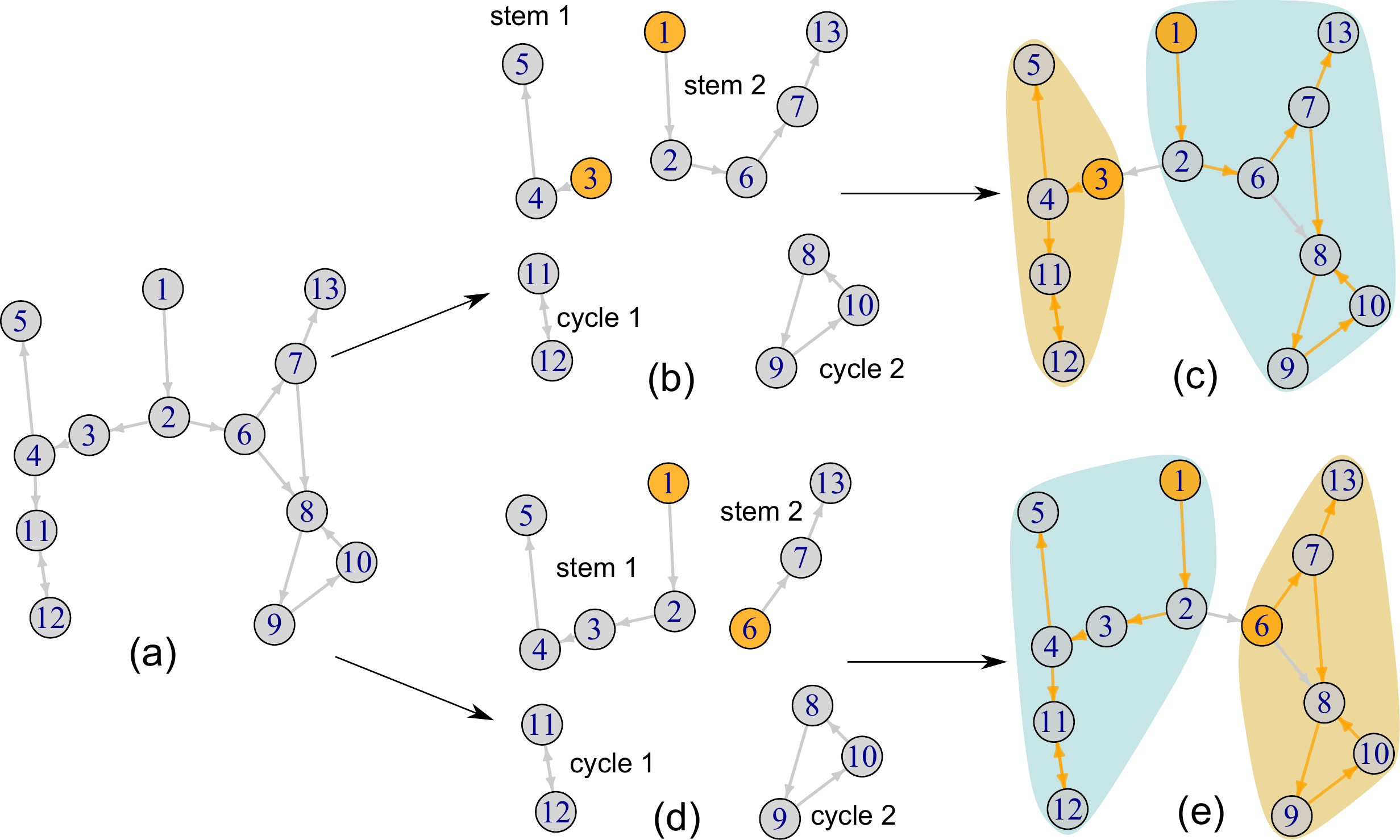}
  \caption{
    (Color online) Controlling a network with minimum driver node sets. (a) a toy network with 13 nodes. (b) One maximum matching with unmatched nodes {1,3}. (c) The cactus  structure corresponding to (b) with driver nodes {1,3}. (d) One maximum matching with unmatched nodes {1,6}. (e) The cactus  structure corresponding to (d) with driver nodes {1,6}. 
    In each cactus structure, orange nodes and grey nodes are driver nodes and non-driver nodes. The subnetwork each driver node controls is in shaded area. }
	\label{p1:figure0}
\end{figure}

Based on structural controllability, \citet{Liu2011} combined tools from network theory and statistical physics to identify minimum sets of driver nodes that allow to control the whole network of $N$ nodes.
Their approach identifies drivers based on \emph{maximum matching}, which denotes the largest set of directed links without common nodes.
That means, this set contains only node-disjoint directed paths, and directed cycles.
In a maximum matching, a node is unmatched if no link in the maximum matching points at it. 
These unmatched nodes form one minimum set of driver nodes. 
In the example of Figure~\ref{p1:figure0}, which we explain further below, $(b)$ and $(d)$ correspond to two configurations of maximum matching. 
Additionally, if we attach a control signal to the top node in a path, we have a stem (see Figure \ref{p1:figure0} b, d). 
If we add the minimum set of links that connects each cycle to only one of the stems, we have a \emph{cactus structure} spanning the network (see Figure \ref{p1:figure0} c, e). 
This cactus structure maintains the controllability of the whole network, with the unmatched nodes as one minimum set of drivers \cite{Lin1974}.  

For an arbitrary network of size $N$, there are multiple sets of driver nodes.
If $N_{c}$ denotes the number of driver nodes, then $\mathrm{MDS}$ denotes the \emph{minimum set} of driver nodes that is required to control the \emph{full network}. 
$N_{d}$ is the size of the set $\mathrm{MDS}$, it can be larger or smaller than $N_{c}$. 
Again, there can be multiple minimum sets of driver nodes. 
We then define the fraction $n_{d}=N_{d}/N$ as the relative size of the minimum set of drivers.
Each of these sets can guarantee controllability of the whole network but does not always contain the same nodes. 
Some nodes appear in every driver node set, while others are redundant and seldom become a driver.
We define the probability of node $i$ to be part of a minimum set of driver nodes as $P(D_{i})$.
This probability is called \emph{control capacity} $\mathcal{K}$ in the literature \citep{Jia2013f}. 

Further, for one minimum set of driver nodes, each driver $i$  controls a non-overlapping subnetwork of size $N_{i}$, which can be identified based on the corresponding cactus structure.
Dependent on which cactus structure is obtained, $N_{i}$ can vary and it's distribution is $f(N_{i})$.
This allows us to define the average $\mean{N_{i}}$ over all minimum sets of drivers in which node $i$ is a driver.

We eventually define the probability that a given node is part of the subnetwork of size $N_{i}$ as $P(N_{i})$.
To illustrate this, we look at the toy network presented in Figure \ref{p1:figure0}(a). 
For this network, there are two minimum sets of driver nodes of size $N_{d}=2$, i.e. there are only two drivers shown in Figure \ref{p1:figure0}(b-c). 
The subnetwork each driver node controls is highlighted with the shaded area. 
Obviously, in each cactus configuration, every node appears only in one of the subnetworks controlled by a driver node. 
In this toy example, we observe that node $1$ appears in both sets of driver nodes, and the size of the subnetwork it controls is larger than that of the other two driver nodes.

In the following, we combine the two information about $P(D_{i})$ and $P(N_{i})$ to define our measure \emph{control contribution}, $\mathcal{C}_{i}$.
Let $\mathrm{MDS}_i$ denote the set of all driver node sets which include node $i$.
The probability that node $i$ becomes a driver node in $\mathrm{MDS}$ is $P(D_i) = {|\mathrm{MDS}_i| }/{|\mathrm{MDS}|}$.
The conditional probability that one node appears in the subnetwork controlled by $i$ given that $i$ is a driver is expressed by
$P(N_i|D_i)=\mean{N_{i}}/N$.
Based on Bayes' rule, we finally calculate the joint probability:
\begin{equation}
P(D_i \cap N_i)=P(N_i| D_i)*P(D_i)
\label{eq:1}
\end{equation}
To calculate $P(D_i \cap N_i)$, it is required to know all possible cactus structures of the network, which is computationally prohibitive for large networks because it can increase with $N!$
Therefore, instead of calculating the precise value of $P(N_i|D_i)$, we restrict ourselves to to identifying its \emph{upper bound},
\begin{align}
  P(N_i|D_i) & \leq \arg\!\max_{ m\in \mathrm{MDS}_{i}}\left({{N_{i}}(m) \over N}\right)=P(E_i)
  \label{eq:2}
\end{align}
This upper bound is the \emph{control range}, $\mathcal{R}$, normalized by the system size.
It gives the maximum size of the subnetwork controlled by node $i$. 
Based on this, we now define the most central measure of our study as
\begin{align}
  \mathcal{C}_{i}=P(E_i)*P(D_i)= \mathcal{R}_{i} \mathcal{K}_{i}
  \label{eq:3}
\end{align}
We denote this as the \emph{control contribution} $\mathcal{C}_{i}$ of node $i$.

\subsection{Calculating control contributions of driver nodes}
\label{sec:calc-contr-contr}

To calculate control contribution $\mathcal{C}$, let us first look at the the toy network presented in Figure \ref{p1:figure0}(a). 
There are only two unique cactus structures with three driver nodes in total: node $1$, $3$ and $6$.
We remind that control range $\mathcal{R}$ is the number of nodes controlled by a given node, normalized to the system size and control capacity $\mathcal{K}$ is the probability for a node to become a driver node. 

With reference to the two cactus structures, we have $\mathcal{R}_{1}=1$, $\mathcal{R}_{3}=0.5$, $\mathcal{R}_{6}=0.5$, and $\mathcal{K}_{1}=8/13$, $\mathcal{K}_{2}=5/13$, $\mathcal{K}_{3}=6/13$. The control contribution for these three driver nodes are: $\mathcal{C}_{1}=8/13$, $\mathcal{C}_{2}=5/26$, $\mathcal{C}_{3}=3/13$. $\mathcal{C}_{1}$ is the highest among all three driver nodes.

For an arbitrary network, we can calculate the control contribution using a random sampling approach. 
This takes into account the fact that, in different samples of cactus structures, the controllable subnetwork of a driver node can be the same.
After accumulating a sufficient number of samples, new cactus structures can hardly produce new controllable subnetworks of larger size for any potential driver node. This property allows us to efficiently approximate control contribution $\mathcal{C}$ using random samples of cactus structures. 

Starting from an initial configuration for the minimum driver set, we generate random samples of cactus structures based on random samples of minimum sets of drivers, as indicated in \cite{Jia2013f}.
By definition, for a minimum set of drivers, each driver node $i$ is the top node in a stem.
The corresponding cactus structure controlled by node $i$ can be constructed by adding cycles that are attached to the stem via a directed link. 

For example, Figure \ref{p1:figure0}(b) shows one maximum matching for the toy network in (a).
This maximum matching contains two stems and two cycles. 
For the driver node $3$, the corresponding stem is stem 1.
It contains the nodes $3$, $4$, $5$, in which  node $3$ is the top node colored in orange.
To construct the cactus structure, we now need to connect the cycles to the stems.
In the toy network (a) we see that cycle $1$ is pointed to by node $4$.
Therefore, we add cycle $1$ to the subnetwork controlled by node $3$.
On the other hand, because there is no link from node $4$ and node $5$ to cycle $2$, we cannot add this cycle to the subnetwork controlled by node 3.
This results in the cactus structure shown in Figure \ref{p1:figure0}(c). 

Note that in an arbitrary network, there can be more than minimum set of drivers that lead to different cactus structures.  
For example, a second configuration of the cactus structure can be seen in Figure \ref{p1:figure0}(d,e). 
Because we limit ourselves to the upper bound of $N_i$, we focus on the cactus structure that identifies the largest controllable subnetwork for each driver node.
In our example, for driver node 1 we only consider the subnetwork shown in (c).
Repeating this construction process for different minimum sets of drivers, we have random samples of cactus structures.

Now, we approximate the control contribution $\mathcal{C}$ based on these samples.
Concretely, we denote $Q$ as the sum of the control contributions of all nodes,  $Q(t)= \sum_{i} \mathcal{C}_i(t)$.
$t$ refers to the number of iterations because the cactus structures are identified in an optimization process.
For each iteration, we identify one cactus structure. 
Then, the quality function which measures the relative increase in $Q$ at each step $t$ can be defined as $\Delta(t)=(Q(t)-Q(t-1))/Q(t-1)$.
$\Delta(t)=0$ indicates that the cactus structure generated at iteration $t$ does not contain larger controllable subnetworks for each driver and therefore will not change with further iterations. 
As shown in the \textbf{Algorithm \ref{algo}} below, if the condition $Q(t)$=0 continues to hold for enough iteration steps $\delta$, this implies we have already generated sufficient samples of cactus structures to approximate the control contribution. The algorithm works as follows:

\begin{algorithm}[H]
	\SetAlgoLined
	Initialization\;
	$Q(0)=0$ \\
	$t=1$ \\
	$marker=0$ \\
	
	\While{$marker < \delta$}{
		produce a random cactus structure \\
		calculate Q(t) and $\Delta(t)$ \\
		\eIf{$\Delta(t)==0$}{
			$marker=marker+1$
		}{
			$marker=0$
		}
		$t=t+1$ \\
	}
	\caption{Random sampling procedure to calculate control contribution.}
	\label{algo}
\end{algorithm}

\section{Results}
\label{sec:results}

\subsection{Distribution of control contribution}
\label{sec:distr-contr-contr}

\begin{table}[htbp]
\centering
\caption{Statistics of networks analyzed in this paper: network size $N$, number of links $E$, average degree $k$, degree correlation $r$, and clustering coefficient $c$.}
\label{table1}
\begin{tabular}{llrd{3.2}d{3.2}d{3.2}}
\toprule
Name                  & $N$    & $L$     & $k$  & $r$   & $c$     \\
\midrule
\rowcolor[HTML]{EFEFEF} 
Erd\H{o}s-R\'enyi & 1000 & 1500  & 3.0    & 0.021  & 0.004 \\
Scale-Free            & 1000 & 4000  & 8.0    & -0.030 & 0.017 \\
\rowcolor[HTML]{EFEFEF} 
SmaGri citation       & 1024 & 4918  & 9.6  & -0.18  & 0.094 \\
Ownership      & 1318 & 12184 & 18.4 & 0.13   & 0.032 \\
\bottomrule 
\end{tabular} 
\end{table}

\paragraph{Data sets.}

To explore the properties of our measure $\mathcal{C}_{i}$,
we consider both synthetic networks and empirical networks as summarized in Table \ref{table1}.
The synthetic networks represent two important classes of network topologies, namely random networks with (a) a (narrow)  Poissonian degree distribution and (b) a (broad) power law degree distribution. 
The empirical networks include the \texttt{SmaGri citation} network, obtained from the \texttt{Pajek} dataset.
It shows how papers cite each other according to the \texttt{WebofScience}.
The second empirical network is an ownership network constructed from the \texttt{ORBIS} 2007 dataset  \cite{Vitali2011a,Zhang2019}.
It contains information about millions of firms and their ownership relations. 
Therefore, in our analysis we restrict ourselves only to the strongly connected component, which contains 1318 firms.
The ownership network is a dense network, while  the citation network is sparse.     
In accordance with the requirement of structural controllability, both the empirical and the synthetic networks are directed.

\paragraph{Distribution of $\mathcal{C}$. }

\begin{figure}[t]
  \centering
  \includegraphics[width=0.95\columnwidth]{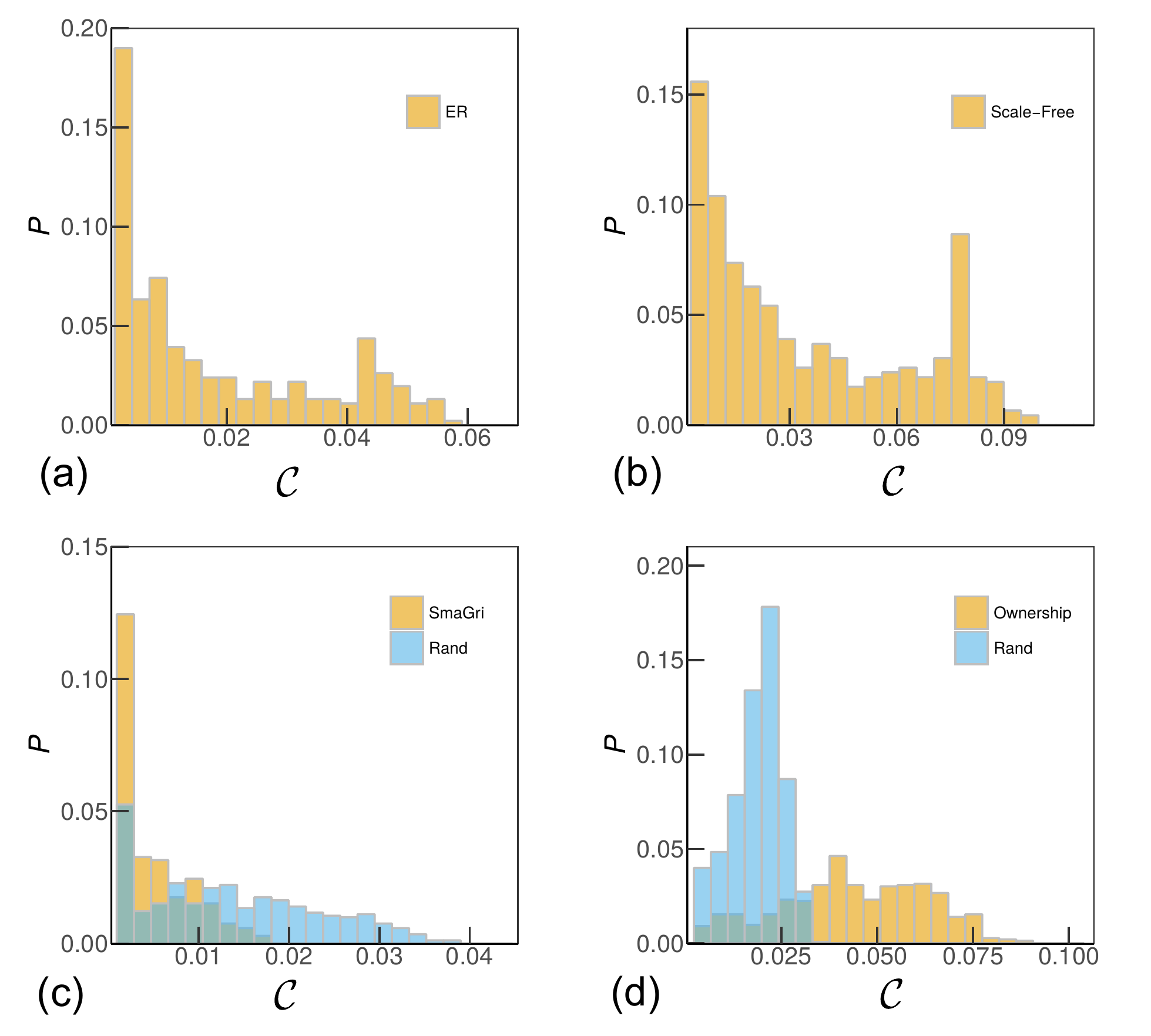}
  \caption{(Color online) Distribution of control contribution $\mathcal{C}$ for driver nodes in real and synthetic networks. In (c-d), we also show the distribution for the randomized version of the network preserving the degree sequence.}
  \label{p1:figure1}
\end{figure}

Figure \ref{p1:figure1} shows the distributions of our measure  $\mathcal{C}$ for the four networks studied. 
We find that, for all networks, the maximum value of $\mathcal{C}$ is less than $0.1$.
This indicates that driver nodes can at most control a small part of the network.

Looking at the form of the distributions, we find that 
for the Erd\H{o}s-R\'enyi network, the scale-free network and the citation network, the distribution of $\mathcal{C}$ is skewed to the left with its peak near zero.
This differs from the distribution for the ownership network, where $\mathcal{C}$ has a peak at the center.
This can be partly attributed to the fact that the ownership network is very dense, as it only considers the strongly connected component. 

To reflect the impact of the particular topology, we compare the two empirical networks with randomly generated networks that have the same degree distribution.
As we observe in the bottom panel of Figure \ref{p1:figure1}, even though the degree distributions are the same, the distributions of $\mathcal{C}$ change significantly, both with respect to the maximum value of $\mathcal{C}$ and the position of the peak. 
These differences indicate that the degree distribution alone is not sufficient to determine the distribution of control contribution.
It is important to note that this observation is different from previous results \citep{Jia2013f,Liu2012} in which the distributions of controllability-based measures, such as control range and control capacity, are mainly determined by the degree distribution.

\subsection{Identifying top driver nodes}
\label{sec:ident-top-driv}

We further compare control contribution $\mathcal{C}$ to other measures that are used to identify top drivers.
Let us recap here that top driver nodes are nodes that together can control a larger part of the network.
Precisely, given a small set of $N_c$ drivers, the size $N_b$ of the  controlled subnetwork becomes as large as possible because these are the top driver nodes. 

In our comparison, we consider six measures in total: three control-based measures, control contribution $\mathcal{C}$, control range $\mathcal{R}$,  control capacity $\mathcal{K}$, two degree-based measures, in-degree and out-degree, and one measure based on a completely random sampling of driver nodes, as a reference case. 
Each of these measures, except for the random case, provides us with a ranking of the nodes.  
From this ranking, we  deterministically choose the top $N_c$ nodes.
Here we have to consider that there are differences between degree-based measures and control-based measures. 
Driver nodes are more likely to be low-degree nodes \cite{Jia2013f,Liu2011}. 
Therefore, if we rank nodes for degree-based measures, we have to rank them in \emph{increasing} order of in(out)-degree, instead of decreasing order, which is done for all control-based measures.

With this set of top driver nodes for the six different measures, we calculate the size of the subnetwork $N_b$  that can be controlled.
Based on this approach, the best measure to identify top driver nodes is the one that leads to the largest controllable system size $N_b$ among all the measures.

Before we come to the results, we emphasize that calculating control range $\mathcal{R}$ and control capacity $\mathcal{K}$ is computationally expensive because it is necessary to generate random samples of cactus structures.
Degree-based measures instead can be easily calculated.

To facilitate the comparison, we focus on the relative size of the controllable subnetwork, i.e., we calculate $n_b=N_b/N$ for each ranking scheme given a fraction of driver nodes $n_c=N_c/N$.
For a given set of drivers, the relative size of the controllable subnetwork $n_b$ can be efficiently determined via linear programming, as shown in \cite{Poljak1990,Zhang2016}.

$n_b=1$ implies that we can control the whole network, and $n_c=1$ indicates that we choose all nodes as drivers directly. 
Since there is always one \emph{minimum set} of driver nodes of size $N_d$ that we can use to control the whole network, we set $n_{d}=N_d/N$ as the upper bound for $n_c$.

\begin{figure}[htbp]
  \centering
  \includegraphics[width=0.95\columnwidth]{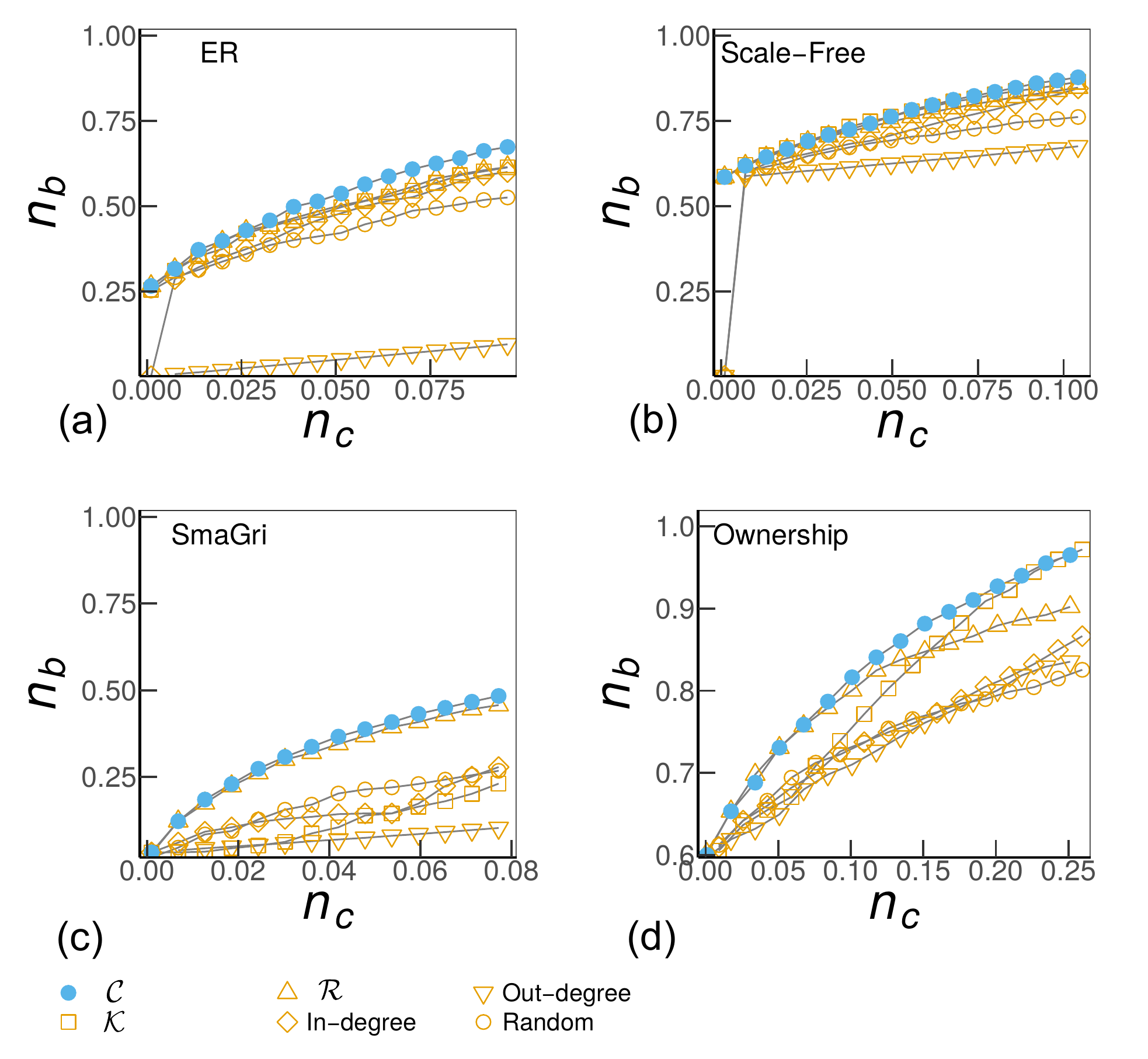}

  \caption{(Color online) $n_b$ as a function of $n_c$ for different ranking schemes.}
  \label{p1:figure4}
\end{figure}

Figure  \ref{p1:figure4} compares $n_b$ for the six ranking schemes as a function of $n_c$. 
Here we highlight two observations: 
(i)  Among the six measures that we use to rank driver nodes, control contribution $\mathcal{C}$ leads to the largest $n_b$, followed by control capacity $\mathcal{K}$.  In comparison, out-degree leads to the smallest $n_b$.
We further discuss this in the next section. 
(ii) Whichever measures we use to rank driver nodes, $n_b$ never reaches 1. This is expected because of the condition $n_c < n_d$.

In conclusion, control contribution $\mathcal{C}$ outperforms the other tested measures when identifying top driver nodes, because nodes chosen according to their $\mathcal{C}_{i}$ values always control a larger part of the network.  
We are aware that the above results are obtained for the particular set of networks used in our study.
However, we have also checked the robustness of our findings with an ensemble approach presented in the supporting information.

\subsection{Control contribution and other structural properties}
\label{sec:contr-contr-other}

So far, we have observed that driver nodes chosen according to their $\mathcal{C}_{i}$ values lead to the largest $n_b$. Now we explore why this is the case. 
We remind that $\mathcal{C}$ is composed of control range $\mathcal{R}$ and control capacity $\mathcal{K}$, according to Eq. \eqref{eq:3}.
Therefore we limit our exploration to these two measures.

\begin{figure}[htbp]
	\centering
	\includegraphics[width=0.7\columnwidth]{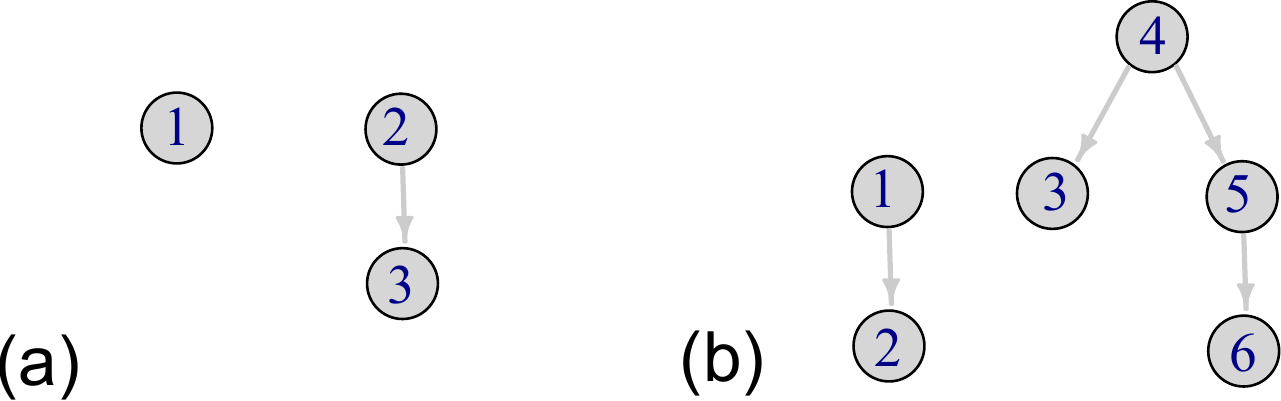}

	\caption{Toy examples to illustrate why control capacity and control range should be considered together in choosing driver nodes.}
	\label{p1:tt}
\end{figure}

We start from two toy examples,  shown in Figure \ref{p1:tt}.
Example (a) contains 3 nodes in which node $1$ is isolated. Based on structural controllability theory, we know that this node is always a driver node,  because it has to be included into the set of driver nodes if one wants to control the \emph{whole} network.
Such an isolated node always has the highest value of control capacity $\mathcal{K}$. 
However, with respect to our performance measure, $N_{b}$, we do not gain a lot from this isolated node, because it only controls itself, which increases $N_b$ by one.
In comparison, node $2$ can control both itself and node $3$ and it is always a driver node, therefore it should be the top driver node in this network. 
This makes clear why control capacity $\mathcal{K}$ alone is a bad measure for top driver nodes.

In example (b),  there are two minimum sets of drivers. One set contains node 1, 4 and 3, the other set contains node 1, 4 and 5. Node $5$ controls both itself and node 6. Therefore, it has a control range $\mathcal{R}$ of $2$. 
However, because it appears in only one of two minimum sets of drivers, it has a low control capacity value $\mathcal{K}$  of $0.5$, and the subnetwork controlled by it can be fully controlled by node $4$ as well. 
This is different from node $1$ which has the same control range but a higher control capacity of $1$.  Therefore, node $1$ should be the top driver in the network.
This example indicates that $\mathcal{R}$ alone is also a bad measure for top driver nodes.

\begin{figure}[htbp]
  \centering
  \includegraphics[width=0.95\columnwidth]{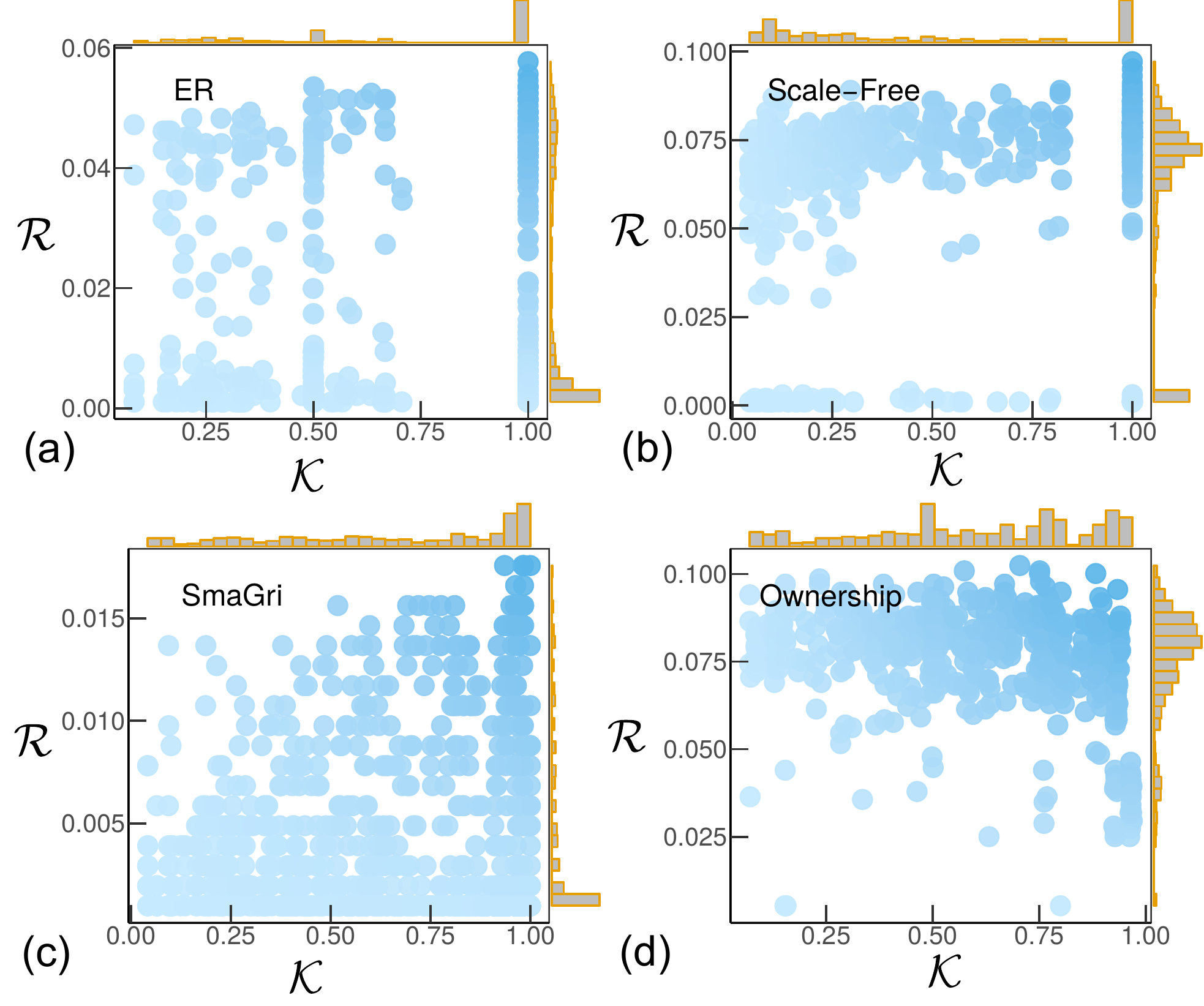}

  \caption{(Color online) Scatter plot to illustrate the absence of strong positive correlation between control capacity and control range. Here we color each point according to the control contribution of the corresponding driver node.}
  \label{p1:figure3}
\end{figure}

The above two examples demonstrate that, in order to identify top driver nodes, we should consider both control capacity $\mathcal{K}$ and control range $\mathcal{R}$.
However, one could still argue that for an arbitrary network, because both  $\mathcal{K}$ and $\mathcal{R}$ are calculated based on random samples of cactus structure, a strong positive correlation could be observed.
If this was the case, there would be no need for a new measure. 
To investigate this conjecture, we explore the correlation of  control capacity $\mathcal{K}$ and control range $\mathcal{R}$  with the scatter plot of driver nodes shown in Figure \ref{p1:figure3}.

We observe that, for all four networks, the values of $\mathcal{R}(\mathcal{K})$  are very broadly scattered.  
Importantly, there are no uniform patterns or strong correlation between control range $\mathcal{R}$ and control capacity $\mathcal{R}$. 
This means that we can hardly predict one measure based on the other measure, and both measures capture individual aspects of drivers. 
This also justifies the motivation for our proposed measure, which combines the two different aspects and provides new information that cannot be covered by either of them, alone.

\section{Discussion}
\label{sec:discussion}

Network controllability helps us to identify the minimum set of driver nodes, $\mathrm{MDS}$, needed to control the whole network.
Under practical circumstances, however, we may not have access to all of these driver nodes or do not want to control the whole network.
Then, the question arises how to choose a smaller set of driver nodes such that, given this number, the largest possible subset of the network can be controlled.
If we have to restrict to this smaller set, we should have a ranking of driver nodes that allows us to pick those that have the largest impact on controlling the network. 

Existing measures for such a ranking, e.g. control capacity, $\mathcal{K}$, and control range, $\mathcal{R}$, are not best suited because they only focus on one aspect of driver nodes, either their probability to become a driver or the size of the subnetwork they control.
As one contribution of this paper, we provide a new measure, \emph{control contribution} $\mathcal{C}$, that combines these two aspects.
We demonstrate that driver nodes chosen according to their $\mathcal{C}$ values always perform better in controlling the network.

As a second contribution, we verify that $\mathcal{C}$ indeed contains information that cannot be traced back to the degree, control capacity or control range of a node.
This was shown both by studying the correlations between these measures and by means of arguments related to the network topology (see Sect.~\ref{sec:contr-contr-other}).

In conclusion, using control contribution $\mathcal{C}$ allows us to identify the driver nodes with the most impact in controlling the network, without the pain of a  'brute force' approach to cope with combinatorial explosion.

\subsection*{Acknowledgements}
All authors designed and performed the research, and wrote the paper. Y.Z. performed all the simulations.

\begin{appendix}
\section{Appendix: Significance in the difference of controllable subspace $N_b$}

As Figure~\ref{p1:figure4} indicates, some of the curves are very close.
Therefore, we test whether the difference in $n_b$ are  significant.
For this, we use an ensemble approach based on 100 synthetic networks with the same network configuration parameters. 
Concretely,
we calculate the areas $S_{\mathcal{C}}$ , $S_{\mathcal{K}}$ , $S_{\mathcal{R}}$  under the respective curves of $n_b$.
To facilitate the comparison, we define two measures, $RS_0=S_{\mathcal{C}}/S_{\mathcal{K}}-1$, and $RS_1=S_{\mathcal{C}}/S_{\mathcal{R}}-1$ to capture the difference in the area sizes.
Obviously, if the measure $\mathcal{C}$ for choosing driver nodes outperforms $\mathcal{K}$ and $\mathcal{R}$, then 
$RS_0$ and $RS_1$ should be larger than $0$.

\begin{figure}[htbp]
  \centering
  \includegraphics[width=0.95\columnwidth]{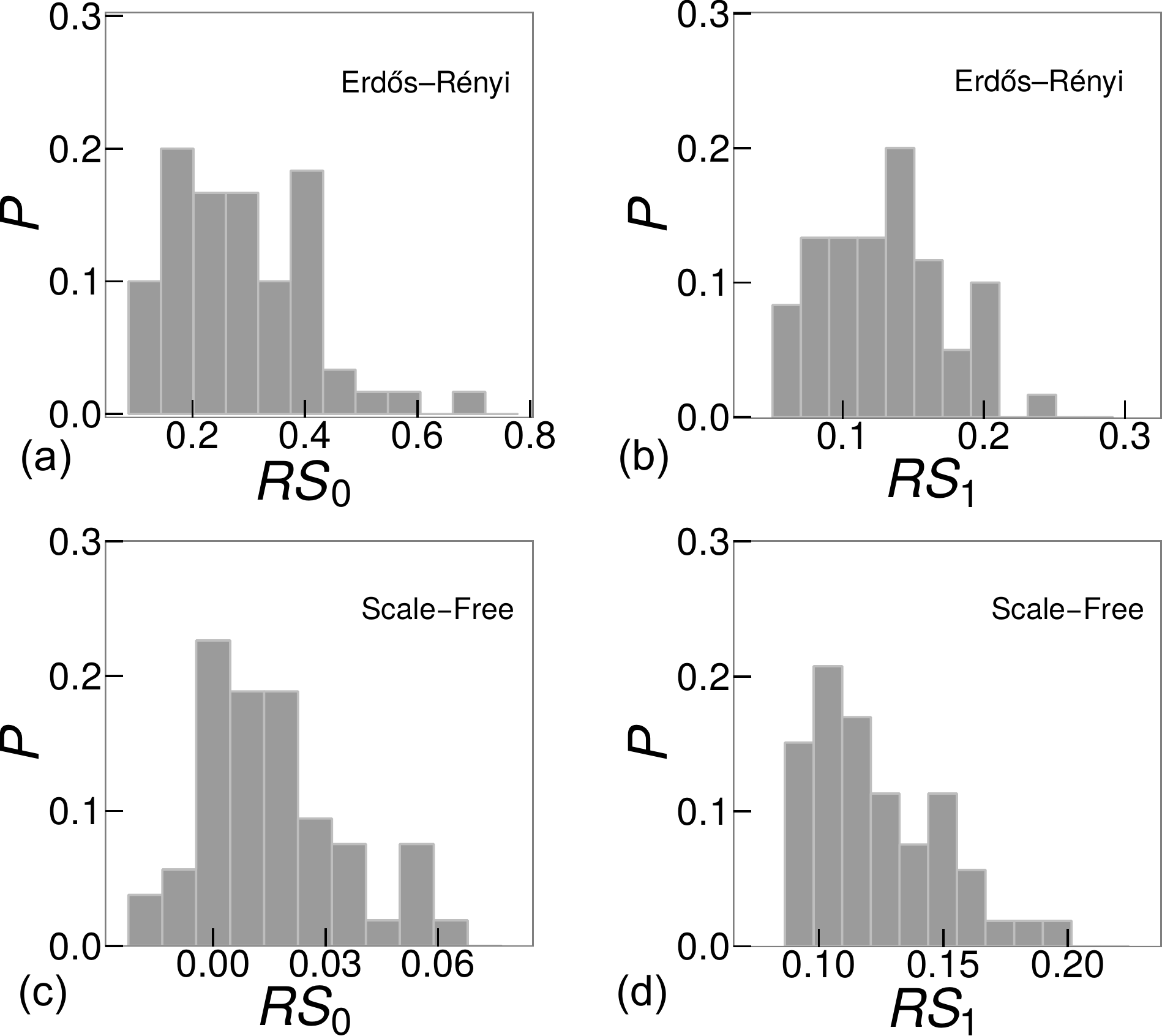}

  \caption{ Distributions of $RS_0$ and $RS_1$ for (a-b) Erd\H{o}s-R\'enyi networks, (c-d) Scale-Free networks. Note the changes of the $x$ scale.}
  \label{p1:figure5}
\end{figure}

Figure \ref{p1:figure5} displays the distribution of $RS_0$ and $RS_1$ obtained from the 100 synthetic networks, both for  Erd\H{o}s-R\'enyi and scale-free networks.
$RS_1$ is always positive, i.e. choosing driver nodes according to control contribution always leads to a larger controllable subnetwork, in comparison to control range. 
For $RS_0$, the major part of the distribution is above zero,  with an associated $P$ value of $0.0001$ at the 95\% confidence interval.  
Therefore, on average, choosing driver nodes with respect to control contribution leads to a larger controllable subnetwork, in comparison to control capacity. 
This means that based on our simulations, $\mathcal{C}$ is the best measure to rank driver nodes compared with existing control-based measures.

\end{appendix}


\begin{thebibliography}{22}
\expandafter\ifx\csname natexlab\endcsname\relax\def\natexlab#1{#1}\fi
\expandafter\ifx\csname url\endcsname\relax
  \def\url#1{\texttt{#1}}\fi
\expandafter\ifx\csname urlprefix\endcsname\relax\def\urlprefix{URL }\fi
\expandafter\ifx\csname selectlanguage\endcsname\relax
  \def\selectlanguage#1{\relax}\fi

\bibitem[{Freeman(1977)}]{Freeman1977}
Freeman, L.~C. (1977).
\newblock {A set of measures of centrality based on betweenness}.
\newblock \emph{Sociometry} \textbf{40}, 35--41.

\bibitem[{Garas \emph{et~al.}(2012)Garas, Schweitzer and Havlin}]{Garas2012b}
Garas, A.; Schweitzer, F.; Havlin, S. (2012).
\newblock {A $\kappa$-shell decomposition method for weighted networks}.
\newblock \emph{New Journal of Physics} \textbf{14}, 083030.

\bibitem[{Jia and Barab{\'{a}}si(2013)}]{Jia2013f}
Jia, T.; Barab{\'{a}}si, A.-L. (2013).
\newblock {Control capacity and a random sampling method in exploring
  controllability of complex networks.}
\newblock \emph{Scientific Reports} \textbf{3}, 2354.

\bibitem[{Kalman(1963)}]{kalman1963}
Kalman, R.~E. (1963).
\newblock {Mathematical Description of Linear Dynamical Systems}.
\newblock \emph{J.S.I.A.M. Control} \textbf{1(2)}, 152--192.

\bibitem[{Kitsak \emph{et~al.}(2010)Kitsak, Gallos, Havlin, Liljeros, Muchnik,
  Stanley and Makse}]{Kitsak2010}
Kitsak, M.; Gallos, L.~K.; Havlin, S.; Liljeros, F.; Muchnik, L.; Stanley,
  H.~E.; Makse, H.~A. (2010).
\newblock {Identification of influential spreaders in complex networks}.
\newblock \emph{Nature Physics} \textbf{6}, 888.

\bibitem[{Lin(1974)}]{Lin1974}
Lin, C.~T. (1974).
\newblock {Structural Controllability}.
\newblock \emph{IEEE Transactions on Automatic Control} \textbf{19(3)},
  201--208.

\bibitem[{Liu and Pan(2014)}]{Liu2014b}
Liu, X.; Pan, L. (2014).
\newblock {Detection of driver metabolites in the human liver metabolic network
  using structural controllability analysis.}
\newblock \emph{BMC Systems Biology} \textbf{8}, 51.

\bibitem[{Liu and Barab{\'{a}}si(2016)}]{Liu2016}
Liu, Y.-Y.; Barab{\'{a}}si, A.-L. (2016).
\newblock {Control principles of complex systems}.
\newblock \emph{Review of Modern Physics} \textbf{88(3)}, 35006.

\bibitem[{Liu \emph{et~al.}(2011)Liu, Slotine and Barabasi}]{Liu2011}
Liu, Y.-Y.; Slotine, J.-J.; Barabasi, A.-L. (2011).
\newblock {Controllability of complex networks}.
\newblock \emph{Nature} \textbf{473(7346)}, 167--173.

\bibitem[{Liu \emph{et~al.}(2012)Liu, Slotine and Barab{\'{a}}si}]{Liu2012}
Liu, Y.-Y.; Slotine, J.-J.; Barab{\'{a}}si, A.-L. (2012).
\newblock {Control centrality and hierarchical structure in complex networks}.
\newblock \emph{PLoS ONE} \textbf{7}, e44459.

\bibitem[{Luenberger(1979)}]{Luenberger1979}
Luenberger, D.~G. (1979).
\newblock \emph{{Introduction to Dynamic Systems Theory, Models, and
  Applications}}.
\newblock Wiley.

\bibitem[{Nacher and Akutsu(2013)}]{Nacher2013c}
Nacher, J.~C.; Akutsu, T. (2013).
\newblock {Structural controllability of unidirectional bipartite networks.}
\newblock \emph{Scientific Reports} \textbf{3}, 1647.

\bibitem[{Poljak(1990)}]{Poljak1990}
Poljak, S. (1990).
\newblock {On the generic dimension of controllable Subspaces}.
\newblock \emph{IEEE Transactions on Automatic Control} \textbf{35}, 367--369.

\bibitem[{Srihari \emph{et~al.}(2013)Srihari, Raman, Leong and
  Ragan}]{Srihari2013}
Srihari, S.; Raman, V.; Leong, H.~W.; Ragan, M.~a. (2013).
\newblock {Evolution and Controllability of Cancer Networks: A Boolean
  Perspective.}
\newblock \emph{IEEE/ACM Transactions on Computational Biology and
  Bioinformatics} \textbf{6}, 83--94.

\bibitem[{Vinayagam \emph{et~al.}(2016)Vinayagam, Gibson, Lee, Yilmazel,
  Roesel, Hu, Kwon, Sharma, Liu, Perrimon and Barab{\'{a}}si}]{Vinayagam2016}
Vinayagam, A.; Gibson, T.~E.; Lee, H.-J.; Yilmazel, B.; Roesel, C.; Hu, Y.;
  Kwon, Y.; Sharma, A.; Liu, Y.-Y.; Perrimon, N.; Barab{\'{a}}si, A.-L. (2016).
\newblock {Controllability analysis of the directed human protein interaction
  network identifies disease genes and drug targets.}
\newblock \emph{Proceedings of the National Academy of Sciences of the United
  States of America} \textbf{113}, 1603992113.

\bibitem[{Vitali \emph{et~al.}(2011)Vitali, Glattfelder and
  Battiston}]{Vitali2011a}
Vitali, S.; Glattfelder, J.; Battiston, S. (2011).
\newblock {The network of global corporate control}.
\newblock \emph{PLoS ONE} \textbf{6(10)}, e25995.

\bibitem[{Wang \emph{et~al.}(2012)Wang, Gao and Gao}]{Wang2012e}
Wang, B.; Gao, L.; Gao, Y. (2012).
\newblock {Control range: a controllability-based index for node significance
  in directed networks}.
\newblock \emph{Journal of Statistical Mechanics} \textbf{2012}, P04011.

\bibitem[{Wuchty(2014)}]{Wuchty2014a}
Wuchty, S. (2014).
\newblock {Controllability in protein interaction networks.}
\newblock \emph{Proceedings of the National Academy of Sciences of the United
  States of America} \textbf{111}, 7156--60.

\bibitem[{{Xiang} \emph{et~al.}(2019){Xiang}, {Chen}, {Ren} and
  {Chen}}]{LX2019}
{Xiang}, L.; {Chen}, F.; {Ren}, W.; {Chen}, G. (2019).
\newblock Advances in Network Controllability.
\newblock \emph{IEEE Circuits and Systems Magazine} \textbf{19(2)}, 8--32.

\bibitem[{Yuan \emph{et~al.}(2014)Yuan, Zhao, Wang, Di and Lai}]{Yuan2014}
Yuan, Z.; Zhao, C.; Wang, W.~X.; Di, Z.; Lai, Y.~C. (2014).
\newblock {Exact controllability of multiplex networks}.
\newblock \emph{New Journal of Physics} \textbf{16}.

\bibitem[{Zhang \emph{et~al.}(2016)Zhang, Garas and Schweitzer}]{Zhang2016}
Zhang, Y.; Garas, A.; Schweitzer, F. (2016).
\newblock {Value of peripheral nodes in controlling multilayer scale-free
  networks}.
\newblock \emph{Physical Review E} \textbf{93(1)}, 1--6.

\bibitem[{Zhang and Schweitzer(2019)}]{Zhang2019}
Zhang, Y.; Schweitzer, F. (2019).
\newblock {The Interdependence of Corporate Reputation and Ownership: A Network
  Approach to Quantify Reputation}.
\newblock \emph{SSRN} \textbf{3337540}.

\end{thebibliography}
\end{document}